\documentclass[12pt,article,onecolumn]{IEEEtran}

\usepackage{srcltx,epsf,amsmath,amssymb,amsfonts,mathrsfs}
\usepackage{times}
\usepackage{subeqnarray}
\usepackage{epsfig}
\usepackage{url}

\newtheorem{theorem}{Theorem}
\newtheorem{lemma}{Lemma}
\newtheorem{remark}{Remark}

\def \md {{\rm\:mod\:} \Lambda}
\def \tm {^{(t)}}
\def \cov {{\rm cov}}
\def \eff {{\rm eff}}
\def \vol {{\rm Vol}}
\def \TE  {{\rm TE}}

\begin{document}

\title{Nested Lattice Codes for Gaussian Relay Networks with Interference}
\author{\authorblockN{Wooseok Nam, \IEEEmembership{Student Member, IEEE,}
        Sae-Young Chung, \IEEEmembership{Senior Member, IEEE,}
        and Yong H. Lee, \IEEEmembership{Senior Member, IEEE}}\\
\authorblockA{School of EECS, KAIST,\\ Daejeon, Republic of Korea\\
E-mail: wsnam@stein.kaist.ac.kr, sychung@ee.kaist.ac.kr,
yohlee@ee.kaist.ac.kr}\thanks{This work was supported by the IT R\&D
program of MKE/IITA. [2008-F-004-01, 5G mobile communication systems
based on beam division multiple access and relays with group
cooperation]} } \maketitle

\begin{abstract}
In this paper, a class of relay networks 
is considered. We assume that, at a node, outgoing channels to its
neighbors are orthogonal, while incoming signals from neighbors can
interfere with each other. We are interested in the multicast
capacity of these networks. As a subclass, we first focus on
Gaussian relay networks with interference and find an achievable
rate using a lattice coding scheme. It is shown that there is a
constant gap between our achievable rate and the information
theoretic cut-set bound. This is similar to the recent result by
Avestimehr, Diggavi,
and Tse, 
who showed such an approximate characterization of the capacity of
general Gaussian relay networks. However, our achievability uses a
structured code instead of a random one. Using the same idea used in
the Gaussian case, we also consider linear finite-field symmetric
networks with interference and characterize the capacity using a
linear coding scheme.
\end{abstract}


\section{Introduction} \label{SEC:Introduction}

Characterizing the capacity of general relay networks has been of
great interest for many years. In this paper, we confine our
interest to the capacity of single source multicast relay networks,
which is still an open problem. For instance, the capacity of single
relay channels is still unknown except for some special cases
\cite{CoverIT79}. However, if we confine the class of networks
further, there are several cases in which the capacity is
characterized.

Recently, in \cite{AhlswedeIT00}, the multicast capacity of wireline
networks was characterized. The capacity is given by the max-flow
min-cut bound, and the key ingredient to achieve the bound is a new
coding technique called network coding. Starting from this seminal
work, many efforts have been made to incorporate wireless effects in
the network model, such as broadcast, interference, and noise. In
\cite{Aref}, the broadcast nature was incorporated into the network
model by requiring each relay node to send the same signal on all
outgoing channels, and the unicast capacity was determined. However,
the model assumed that the network is deterministic (noiseless) and
has no interference in reception at each node. In
\cite{RatnakarIT06}, the work was extended to multicast capacity. In
\cite{AvestimehrAllerton07}, the interference nature was also
incorporated, and an achievable multicast rate was computed. This
achievable rate has a cut-set-like representation and meets the
information theoretic cut-set bound \cite{CoverText} in some special
cases. To incorporate the noise, erasure networks with broadcast or
interference only were considered in \cite{DanaIT06, SmithITW07}.
However, the network models in \cite{DanaIT06}, \cite{SmithITW07}
assumed that the side information on the location of all erasures in
the network is provided to destination nodes. Noisy networks without
side information at destination nodes were considered in
\cite{NazerET08} and \cite{NamAllerton08} for finite-field additive
noise and erasure cases, respectively.

Along the same lines of the previous work on wireless networks
mentioned above, we consider the multicast problem in a special
class of networks called relay networks with interference. More
specifically, we assume that all outgoing channels at each node are
orthogonal, e.g., using frequency or time division multiplexing, but
signals incoming from multiple neighbor nodes to a node can
interfere with each other. Since wireless networks are often
interference limited, our setup focuses on the more important aspect
of them. This model covers those networks considered in
\cite{SmithITW07, NazerISIT06, NazerAllerton07, NazerET08}. Our
interest in the relay networks with interference was inspired by
\cite{ElGamalIT05}, in which the capacity of single relay channels
with interference was established. In this paper, we focus on two
special subclasses of general networks with interference; Gaussian
relay networks with interference and linear finite-field symmetric
networks with interference.

For the Gaussian relay networks with interference, we propose a
scheme based on nested lattice codes \cite{ErezIT04} which are
formed from a lattice chain and compute an achievable multicast
rate. The basic idea of using lattice codes is to exploit the
structural gain of {\em computation coding} \cite{NazerIT07}, which
corresponds to a kind of combined channel and network coding.
Previously, lattices were used in Gaussian networks in
\cite{NazerAllerton07}, and an achievability was shown. However, our
network model differs from the one in \cite{NazerAllerton07} in that
we assume general unequal power constraints for all incoming signals
at each node, while an equal power constraint was mainly considered
in \cite{NazerAllerton07}. In addition, our lattice scheme is
different from that in \cite{NazerAllerton07} in that we use
lattices to produce nested lattice codes, while lattices were used
as a source code in \cite{NazerAllerton07}.

We also show that our achievable rate is within a constant number of
bits from the information theoretic cut-set bound of the network.
This constant depends only on the network topology and not on other
parameters, e.g., transmit powers and noise variances. This is
similar to the recent result in \cite{AvestimehrISIT08}, which
showed an approximate capacity characterization for general Gaussian
relay networks using a random coding scheme. However, our
achievability uses a structured code instead of a random one. Thus,
our scheme has a practical interest because structured codes may
reduce the complexity of encoding and decoding.

Finally, we introduce a model of linear finite-field symmetric
networks with interference, which generalizes those in
\cite{NazerET08, NamAllerton08}. In the finite-field case, we use a
linear coding scheme, which corresponds to the finite-field
counterpart of the lattice coding scheme. The techniques for
deriving an achievable rate for the finite-field network are
basically the same as those for the Gaussian case. However, in this
case, the achievable rate always meets the information theoretic
cut-set bound, and, thus, the capacity is fully established.

This paper is organized as follows. Section \ref{SEC:SystemModel}
defines notations and parameters used in this paper and introduces
the network model and the problem of interest. In Section
\ref{SEC:Gaussian}, we analyze Gaussian relay networks with
interference and give the upper and lower bounds for the multicast
capacity. In Section \ref{SEC:Finite}, we define a model of linear
finite-field symmetric networks with interference and present the
multicast capacity. Section \ref{SEC:Conclusion} concludes the
paper.

\section{Relay networks with interference} \label{SEC:SystemModel}

\subsection{System model and notations}

\begin{figure} [t]
\begin{center}
\epsfig{file=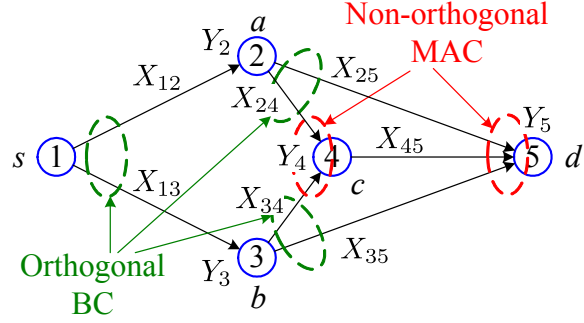, width=3in} \caption{Example of general
memoryless relay network with interference.} \label{FIG:General}
\end{center}
\end{figure}

We begin with a description of the class of networks that will be
dealt with in this paper. The memoryless relay networks with
interference are characterized such that all outgoing channels from
a node to its neighbors are orthogonal to each other. We still
assume that incoming signals at a node can interfere with each other
through a memoryless multiple-access channel (MAC). An example of
this class of networks is shown in Fig. \ref{FIG:General}. Some
special cases and subclasses of these networks have been studied in
many previous works \cite{SmithITW07, NazerISIT06, NazerAllerton07,
NamAllerton08, ElGamalIT05}.

\psfull
\psdraft

We will begin by giving a detailed description of the network and
some definitions of the parameters. The network is represented by a
directed graph $\mathcal{G} = (V,E)$, where $V = \{1, \ldots, |V|\}$
is a vertex set and $E \subseteq V \times V$ is an edge set. Each
vertex and edge correspond to a communication node and a channel in
the network, respectively. In this paper, we focus on a multicast
network: vertex 1 represents the source node and is denoted by $s$,
and the set of destination nodes is denoted by $D$, where $s \notin
D$. It will be assumed that the source node has no incoming edge,
and the destination nodes have no outgoing edge. All the other
nodes, which are neither the source nor the destination, are called
the relay nodes. Since all broadcast channels in the network are
orthogonal, we associate a discrete or continuous random variable
$X_{u,v}\tm$ at time $t$ with edge $(u,v) \in E$ as a channel input
(output of a node). As a channel output (input of a node), we
associate a discrete or continuous random variable $Y_v\tm$ at time
$t$ with node $v \in V \setminus \{1\}$. From now on, we sometimes
drop the superscript `$^{(t)}$' when doing so causes no confusion.

At node $v \in V$, the set of incoming and outgoing nodes are
denoted by
\begin{align*}
\Delta (v) &= \left\{ u: (u,v) \in E \right\} \text, \\
\Theta (v) &= \left\{ w: (v,w) \in E \right\} \text.
\end{align*}
Set $S \subset V$ is called a cut if it contains node $s$ and its
complement $S^c$ contains at least one destination node $d\in D$,
i.e., $S^c \cap D \neq \emptyset$. Let $\Gamma$ denote the set of
all cuts. The boundaries of $S$ and $S^c$ are defined as
\begin{align*}
\bar{S} &= \left\{ u: \exists v \text{ s.t. } (u,v)\in
E, u \in S, v \in S^c \right\}\text,\\
\bar{S}^c &= \left\{ v: \exists u \text{ s.t. } (u,v)\in E, u \in S,
v \in S^c \right\} \text.
\end{align*}
For node $v \in S^c$, the set of incoming nodes across $S$ is
defined as
\begin{equation*}
\Delta_S (v) = \Delta (v) \cap S = \Delta (v) \cap \bar{S} \text.
\end{equation*}
For any sets $S_1 \subseteq V$ and $S_2 \subseteq V$, we define
\begin{align*}
X_{S_1, S_2} &= \left\{ X_{u,v}: (u,v) \in E, u \in S_1, v \in S_2
\right\}
\text, \\
Y_{S_1} &= \left\{ Y_v : v \in S_1 \right\} \text,
\end{align*}
and
\begin{equation*}
X_{\Delta(v)} = \left\{ X_{u,v}: u \in \Delta(v) \right\} \text.
\end{equation*}

Using the aforementioned notations, we can formally define the class
of networks of interest. The memoryless relay network with
interference is characterized by the channel distribution function
\begin{equation*}
p \left(y_V | x_{V,V} \right) = p \left(y_2 | x_{\Delta(2)}\right) p
\left(y_3 | x_{\Delta(3)} \right) \cdots p \left(y_M | x_{\Delta(M)}
\right)
\end{equation*}
over all input and output alphabets.

\subsection{Coding for the relay network with interference}

The multicast over the relay network consists of encoding functions
$f_{u,v}^{(t)}(\cdot)$, $(u,v) \in E$, $t=1,\ldots,N$, and decoding
functions $g_d (\cdot)$, $d \in D$. The source node $s$ has a random
message $W$ that is uniform over $\{1,\ldots,M \}$ and transmits
\begin{equation*}
X_{s,w}^{(t)} = f_{s,w}^{(t)} (W)
\end{equation*}
at time $t$ on the outgoing channels $(s,w)$, $w \in \Theta (s)$.
The relay node $v$ transmits
\begin{equation*}
X_{v,w}^{(t)} = f_{v,w}^{(t)}(Y_v^{t-1})
\end{equation*}
at time $t$ on the outgoing channels $(v,w)$, $w \in \Theta (v)$,
where $Y_v^{t-1} = \left( Y_v^{(1)}, \ldots, Y_v^{(t-1)} \right)$.
At destination node $d \in D$, after time $N$, an estimate of the
source message is computed as
\begin{equation*}
\hat{W} = g_d \left(Y_d^{N}\right) \text.
\end{equation*}
Then, the probability of error is
\begin{equation}
P_e = {\rm Pr} \left\{ \underset{d \in D}{\cup} \left\{ g_d (Y_d^N)
\neq W \right\} \right\} \text. \label{EQ:TotalError}
\end{equation}
We say that the multicast rate $R$ is {\em achievable} if, for any
$\epsilon >0$ and for all sufficiently large $N$, encoders and
decoders with $M \geq 2^{NR}$ exist such that $P_e \leq \epsilon$.
The {\em multicast capacity} is the supremum of the achievable
multicast rates.

As stated in Section \ref{SEC:Introduction}, we are interested in
characterizing the multicast capacity of the memoryless relay
networks with interference. However, as shown in
\cite{NamAllerton08}, even for a relatively simple parallel relay
channel, finding the capacity is not easy. Thus, we further restrict
our interest to the Gaussian networks in Section \ref{SEC:Gaussian}
and the linear finite-field symmetric networks in Section
\ref{SEC:Finite}.

\section{Gaussian relay networks with interference} \label{SEC:Gaussian}

In this section, we consider Gaussian relay networks with
interference. At node $v$ at time $t$, the received signal is given
by
\begin{equation*}
Y_v\tm = \sum_{u \in \Delta(v)} X_{u,v}\tm + Z_v\tm \text,
\end{equation*}
where $Z_v\tm$ is an independent identically distributed (i.i.d.)
Gaussian random variable with zero mean and unit variance. For each
block of channel input $\left( X_{u,v}^{(1)},\ldots,X_{u,v}^{(n)}
\right)$, we have the average power constraint given by
\begin{equation*}
\frac{1}{n} \sum_{t=1}^n \left( X_{u,v}^{(t)} \right)^2 \leq P_{u,v}
\text.
\end{equation*}
In \cite{NazerAllerton07}, Nazer et al. studied the achievable rate
of the Gaussian relay networks with interference for the equal power
constraint case, where $P_{u,v} = P_v$ for all $u \in \Delta(v)$. In
our work, we generalize it such that $P_{u,v}$'s can be different.
The main result of this section is as follows.

\begin{theorem}
For a Gaussian relay network with interference, an upper bound for
the multicast capacity is given by
\begin{equation}
\underset{S\in\Gamma}{\min} \sum_{v \in \bar{S}^c} C\left( \left(
\sum_{u \in \atop \Delta_S (v)} \sqrt{P_{u,v}} \right)^2 \right)
\text, \label{EQ:GaussianUB}
\end{equation}
where $C(x) = \frac{1}{2} \log \left(1+x\right)$. For the same
network, we can achieve all rates up to
\begin{equation}
\underset{S\in\Gamma}{\min} \sum_{v \in \bar{S}^c} \left[
\frac{1}{2} \log \left( \left( \frac{1}{\sum_{u \in \atop \Delta
(v)} P_{u,v}} + 1 \right) \cdot \underset{u \in \atop \Delta_S
(v)}{\max} P_{u,v} \right) \right]^+ \text, \label{EQ:GaussianAch}
\end{equation}
where $[x]^+ \triangleq \max \{x,0\}$. Furthermore, the gap between
the upper bound and the achievable rate is bounded by
\begin{equation}
\sum_{v \in V \setminus \{1\} } \log \left( | \Delta(v) | \right)
\text. \label{EQ:GaussianGap}
\end{equation}
\label{TH:Gaussian}
\end{theorem}

\begin{remark}
Note that, in the equal power case, i.e., $P_{u,v}=P$, the
achievable multicast rate (\ref{EQ:GaussianAch}) has terms in the
form of $\log\left(\frac{1}{K}+P\right)$ for some integer $K \geq
1$. Similar forms of achievable rate were observed in
\cite{NazerAllerton07, NamIZS08, NarayananAllerton07,
PhilosofISIT07} for some equal power Gaussian networks.
\end{remark}

The following subsections are devoted to proving Theorem
\ref{TH:Gaussian}.

\subsection{Upper bound} \label{SEC:GaussianUB}

The cut-set bound \cite{CoverText} of the network is given by
\begin{equation}
R \leq \underset{p( x_{V,V})}{\max} \underset{S \in \Gamma}{\min} \;
I \left( X_{S,V};Y_{S^c} | X_{S^c,V} \right) \text.
\label{EQ:Cutset}
\end{equation}
Though the cut-set bound is a general and convenient upper bound for
the capacity, it is sometimes challenging to compute the exact
cut-set bound in a closed form. This is due to the optimization by
the joint probability density function (pdf) $p( x_{V,V})$. In some
cases, such as the finite-field networks in
\cite{AvestimehrAllerton07, SmithITW07, NazerET08, NamAllerton08},
it is easy to compute the cut-set bound because a product
distribution maximizes it. For the Gaussian case, however, the
optimizing distribution for the cut-set bound is generally not a
product distribution.

Thus, we consider another upper bound which is easier to compute
than the cut-set bound. This bound is referred to as the {\em
relaxed cut-set bound} and given by
\begin{equation}
R \leq \underset{S \in \Gamma}{\min} \underset{p(x_{V,V})}{\max} I
\left( X_{S,V};Y_{S^c} | X_{S^c,V} \right) \text. \label{EQ:Rcutset}
\end{equation}
Due to the max-min inequality, the relaxed cut-set bound is looser
than the original cut-set bound (\ref{EQ:Cutset}). For the relay
network with interference, we can further simplify
(\ref{EQ:Rcutset}) as
\begin{align*}
I (X_{S,V};Y_{S^c}|X_{S^c,V}) &= I(X_{S,S}, X_{S,S^c}; Y_{S^c}|
X_{S^c,V}) \\
&= I(X_{S,S^c}; Y_{S^c}| X_{S^c,V}) \\
&= I(X_{\bar{S},\bar{S}^c}; Y_{\bar{S}^c}| X_{S^c,V}) \text,
\end{align*}
where the second and the third equalities follow by the structure of
the network, i.e.,
\begin{itemize}
\item $X_{S,S} \rightarrow (X_{S,S^c},X_{S^c,V}) \rightarrow Y_{S^c}$,
\item $( X_{S,S^c}, Y_{\bar{S}^c}) \rightarrow X_{S^c,V} \rightarrow
Y_{S^c \setminus \bar{S}^c}$,
\item $X_{S,S^c} = X_{\bar{S},\bar{S}^c}$.
\end{itemize}
For cut $S$, the mutual information $I(X_{\bar{S},\bar{S}^c};
Y_{\bar{S}^c}| X_{S^c,V})$ is maximized when there is a perfect
coherence between all inputs to a Gaussian MAC across the cut. Thus,
we have
\begin{equation}
\underset{p(x_{V,V})}{\max} I(X_{\bar{S},\bar{S}^c}; Y_{\bar{S}^c}|
X_{S^c,V}) = \sum_{v \in \bar{S}^c} C\left( \left( \sum_{u \in \atop
\Delta_S (v)} \sqrt{P_{u,v}} \right)^2 \right) \text.
\label{EQ:Rcutset2}
\end{equation}
Then by (\ref{EQ:Rcutset}) and (\ref{EQ:Rcutset2}), the upper bound
(\ref{EQ:GaussianUB}) follows.

\subsection{Lattices and nested lattice codes}

\psfull
\begin{figure} [t]
\begin{center}
\epsfig{file=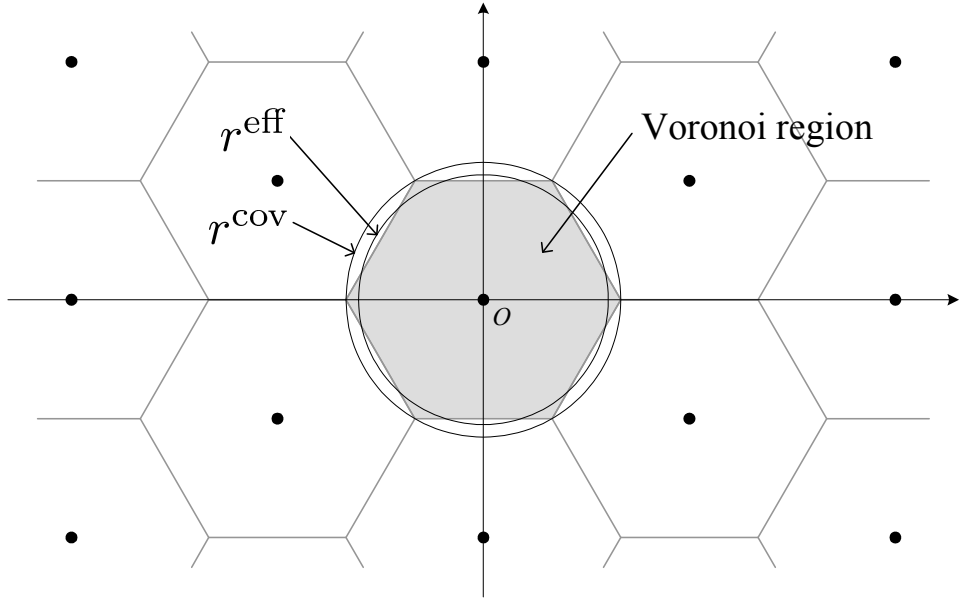, width=5in} \caption{Example:
two-dimensional lattice constellation.} \label{FIG:Lattice}
\end{center}
\end{figure}
\psdraft

Before proving the achievable part of Theorem \ref{TH:Gaussian}, let
us establish some preliminaries for the lattices and nested lattice
codes, which are key ingredients of our achievability proof. For a
more comprehensive review on lattices and nested lattice codes,
see~\cite{ErezIT04, ErezIT05, ForneyAllerton03}. An $n$-dimensional
lattice $\Lambda$ is defined as a discrete subgroup of Euclidean
space $\mathbb{R}^n$ with ordinary vector addition. This implies
that for any lattice points $\lambda, \lambda' \in \Lambda$, we have
$\lambda + \lambda' \in \Lambda$, $\lambda - \lambda' \in \Lambda$,
and ${\bf 0} \in \Lambda$. The nearest neighbor lattice quantizer
associated with $\Lambda$ is defined as
\begin{equation*}
Q({\bf x}) = \underset{\lambda \in \Lambda}{\arg\min}\; \| {\bf x} -
\lambda \| \text,
\end{equation*}
and the $\md$ operation is
\begin{equation*}
{\bf x} \md = {\bf x} - Q({\bf x}) \text.
\end{equation*}
The (fundamental) Voronoi region of $\Lambda$, denoted by
$\mathcal{R}$, is defined as the set of points in $\mathbb{R}^n$
closer to the origin than to any other lattice points, i.e.,
\begin{equation*}
\mathcal{R} = \{ {\bf x} : Q({\bf x}) = {\bf 0} \} \text,
\end{equation*}
where ties are broken arbitrarily. In Fig. \ref{FIG:Lattice}, an
example of a two-dimensional lattice, and its Voronoi region are
depicted.

We now define some important parameters that characterize the
lattice. The covering radius of the lattice $r^{\cov}$ is defined as
the radius of a sphere circumscribing around $\mathcal{R}$, i.e.,
\begin{equation*}
r^{\cov} = \min \;\{ r : \mathcal{R} \subseteq r \mathcal{B} \}
\text,
\end{equation*}
where $\mathcal{B}$ is an $n$-dimensional unit sphere centered at
the origin, and, thus, $r \mathcal{B}$ is a sphere of radius $r$. In
addition, the effective radius of $\Lambda$, denoted by $r^{\eff}$,
is the radius of a sphere with the same volume as $\mathcal{R}$,
i.e.,
\begin{equation*}
r^{\eff} = \left( \frac{\vol(\mathcal{R})}{\vol(\mathcal{B})}
\right)^{\frac{1}{n}} \text,
\end{equation*}
where $\vol(\cdot)$ denotes the volume of a region. The second
moment per dimension of $\Lambda$ is defined as the second moment
per dimension associated with $\mathcal{R}$, which is given by
\begin{equation*}
\sigma^2 (\mathcal{R}) = \frac{1}{\vol(\mathcal{R})} \cdot
\frac{1}{n} \int_{\mathcal{R}} \| {\bf x}\|^2 d{\bf x} \text.
\end{equation*}
In the rest of this paper, we also use $\vol(\Lambda)$ and $\sigma^2
(\Lambda)$, which have the same meaning as $\vol(\mathcal{R})$ and
$\sigma^2 (\mathcal{R})$, respectively. Finally, we define the
normalized second moment of $\Lambda$ as
\begin{equation*}
G(\Lambda) = \frac{\sigma^2 (\mathcal{R})}{\left( \vol(\mathcal{R})
\right)^{2/n}} \text.
\end{equation*}
For any $\Lambda$, $G(\Lambda)$ is greater than $\frac{1}{2 \pi e}$,
which is the normalized second moment of a sphere whose dimension
tends to infinity.

\vspace{2mm}{\bf \em Goodness of lattices}\vspace{2mm}

We consider a sequence of lattices $\Lambda^n$. The sequence of
lattices is said to be {\em Rogers-good} if
\begin{equation*}
\lim_{n \rightarrow \infty} \frac{r^{\cov}}{r^{\eff}} = 1 \text,
\end{equation*}
which implies that $\Lambda^n$ is asymptotically efficient for
sphere covering \cite{ErezIT05}. This also implies the goodness of
$\Lambda^n$ for mean-square error quantization, i.e.,
\begin{equation*}
\lim_{n \rightarrow \infty} G(\Lambda^n) = \frac{1}{2 \pi e} \text.
\end{equation*}
We now define the goodness of lattices related to the channel coding
for the additive white Gaussian noise (AWGN) channel. A sequence of
lattices is said to be {\em Poltyrev-good} if, for $\bar{\bf Z} \sim
\mathcal{N}({\bf 0},\bar{\sigma}^2 {\bf I})$,
\begin{equation}
{\rm Pr} \{ \bar{\bf Z} \notin \mathcal{R} \} \leq e^{-n E_P (\mu)}
\text, \label{EQ:Poltyrev}
\end{equation}
where $E_P(\cdot)$ is the Poltyrev exponent \cite{PoltyrevIT94} and
$\mu$ is the volume-to-noise ratio (VNR) defined as
\begin{equation*}
\mu = \frac{(\vol(\mathcal{R}))^{2/n}}{2 \pi e \bar{\sigma}^2}\text.
\end{equation*}
Note that (\ref{EQ:Poltyrev}) upper bounds the error probability of
the nearest lattice point decoding (or equivalently, Euclidean
lattice decoding) when we use lattice points as codewords for the
AWGN channel. Since $E_P (\mu) > 0$ for $\mu >1$, a necessary
condition for reliable decoding is $\mu >1$.

\vspace{2mm}{\bf \em Nested lattices codes}\vspace{2mm}

Now we consider two lattices $\Lambda$ and $\Lambda_C$. Assume that
$\Lambda$ is coarse compared to $\Lambda_C$ in the sense that
$\vol(\Lambda) \geq \vol(\Lambda_C)$. We say that the coarse lattice
$\Lambda$ is a sublattice of the fine lattice $\Lambda_C$ if
$\Lambda \subseteq \Lambda_C$ and call the quotient group
(equivalently, the set of cosets of $\Lambda$ relative to
$\Lambda_C$) $\Lambda_C / \Lambda$ a {\em lattice partition}. For
the lattice partition, the {\em set of coset leaders} is defined as
\begin{equation*}
\mathcal{C} = \{ \Lambda_C \md \} \triangleq \{ \Lambda_C \cap
\mathcal{R} \} \text,
\end{equation*}
and the {\em partitioning ratio} is
\begin{equation*}
\rho = |\mathcal{C}|^{\frac{1}{n}} = \left(
\frac{\vol(\Lambda)}{\vol(\Lambda_C)} \right)^{\frac{1}{n}} \text.
\end{equation*}

Formally, a lattice code is defined as an intersection of a lattice
(possibly translated) and a bounding (shaping) region, which is
sometimes a sphere. A {\em nested lattice code} is a special class
of lattice codes, whose bounding region is the Voronoi region of a
sublattice. That is, the nested lattice code is defined in terms of
lattice partition $\Lambda_C / \Lambda$, 
in which $\Lambda_C$ is used as codewords
and $\Lambda$ is used for shaping. The coding rate of the nested
lattice code is given by
\begin{equation*}
\frac{1}{n} \log |\mathcal{C}| = \log \rho \text.
\end{equation*}
Nested lattice codes have been studied in many previous articles
\cite{ZamirIT02, ErezIT04, ForneyAllerton03, Krithivasan07}, and
proved to have many useful properties, such as achieving the
capacity of the AWGN channel. In the next subsection, we deal with
the nested lattice codes for the achievability proof of Theorem
\ref{TH:Gaussian}.

\subsection{Nested lattice codes for a Gaussian MAC}
\label{SEC:LatticeCode}

As an achievable scheme, we use a lattice coding scheme.
In~\cite{NazerAllerton07}, lattices were also used to prove an
achievable rate of Gaussian relay networks with interference (called
Gaussian MAC networks). However, they used the lattice as a source
code with a distortion and then related the achievable distortion to
the information flow through the network. Our approach is different
from~\cite{NazerAllerton07} in that we use lattices to produce
coding and shaping lattices, and form nested lattice codes. 
As a result, our approach can handle unequal power constraints where
incoming links have different power at a MAC. Our scheme is a
generalization of the nested lattice codes used for the Gaussian
two-way relay channel in~\cite{NamIZS08, NarayananAllerton07}.

Let us consider a standard model of a Gaussian MAC with $K$ input
nodes:
\begin{equation}
Y = \sum_{j=1}^{K} X_j + Z \text, \label{EQ:StdMAC}
\end{equation}
where $Z$ denotes the AWGN process with zero mean and unit variance.
Each channel input $X_i$ is subject to the average power constraint
$P_i$, i.e., $\frac{1}{n} \sum_{t=1}^n ( X_i^{(t)} )^2 \leq P_i$.
Without loss of generality, we assume that $P_1 \geq P_2 \geq \cdots
\geq P_K$.

The standard MAC in (\ref{EQ:StdMAC}) is a representative of MACs in
the Gaussian relay network with interference. Now, we introduce
encoding and decoding schemes for the standard MAC. Let us first
consider the following theorem which is a key for our code
construction.

\begin{theorem}
For any $P_1 \geq P_2 \geq \cdots \geq P_K \geq 0$ and $\gamma \geq
0$, a sequence of $n$-dimensional lattice chains $\Lambda_1^{n}
\subseteq \Lambda_2^{n}\subseteq \cdots \subseteq
\Lambda_K^{n}\subseteq \Lambda_C^{n}$ exists that satisfies the
following properties.

\noindent a) $\Lambda_i^{n}$, $1 \leq i \leq K$, are simultaneously
Rogers-good and Poltyrev-good while $\Lambda_C^{n}$ is
Poltyrev-good.

\noindent b) For any $\delta > 0$, $P_i - \delta \leq \sigma^2
(\Lambda_i^n) \leq P_i$, $1 \leq i \leq K$, for sufficiently large
$n$.

\noindent c) The coding rate of the nested lattice code associated
with the lattice partition $\Lambda_C^n / \Lambda_K^n$
can approach any value as $n$ tends to infinity, i.e.,
\begin{equation*}
R_K \triangleq \frac{1}{n} \log | \mathcal{C}_K | 
= \gamma + o_n (1) \text,
\end{equation*}
where $\mathcal{C}_K = \left\{ \Lambda_C^n \md_K^n \right\}$ and
$o_n(1) \rightarrow 0$ as $n \rightarrow \infty$. Furthermore, for
$1 \leq i \leq K-1$, the coding rate of the nested lattice code
associated with $\Lambda_C^n / \Lambda_i^n$ is given by
\begin{equation*}
R_i \triangleq \frac{1}{n} \log | \mathcal{C}_i | 
= R_K + \frac{1}{2} \log \left(
\frac{P_i}{P_K} \right) + o_n (1) \text,
\end{equation*}
where $\mathcal{C}_i = \left\{ \Lambda_C^n \md_i^n \right\}$.
\label{TH:LatticeChain}
\end{theorem}

\begin{IEEEproof}
See Appendix \ref{SEC:LatticeChain}.
\end{IEEEproof}

\begin{figure} [t]
\begin{center}
\epsfig{file=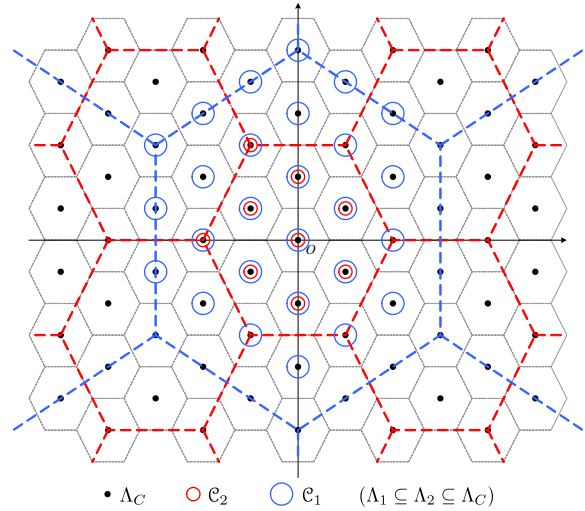, width=3in} \caption{Example of
lattice chain and sets of coset leaders. 
} \label{FIG:LatChain}
\end{center}
\end{figure}

A conceptual representation of the lattice chain and the
corresponding sets of coset leaders are given in Fig.
\ref{FIG:LatChain} for a two-dimensional case.

\psfull
\psdraft

\vspace{2mm}{\bf \em Encoding}\vspace{2mm}

We consider a lattice chain as described in Theorem
\ref{TH:LatticeChain}. We assign the $i$-th input node to the MAC
with the set of coset leaders $\mathcal{C}_i$. For each input node,
the message set $\left\{ 1, \ldots, 2^{nR_i} \right\}$ is
arbitrarily mapped onto $\mathcal{C}_i$. We also define random
dither vectors ${\bf U}_i \sim {\rm Unif}(\mathcal{R}_i)$, $1\leq i
\leq K$, where $\mathcal{R}_i$ denotes the Voronoi region of
$\Lambda_i$ (we dropped the superscript `$^n$' for simplicity).
These dither vectors are independent of each other and also
independent of the message of each node and the noise. We assume
that each ${\bf U}_i$ is known to both the $i$-th input node and the
receiver. To transmit a message that is uniform over $\left\{ 1,
\ldots, 2^{nR_i} \right\}$, node $i$ chooses ${\bf W}_i \in
\mathcal{C}_i$ associated with the message and sends
\begin{equation*}
{\bf X}_i = ( {\bf W}_i + {\bf U}_i ) {\rm\;mod\;} \Lambda_i \text.
\end{equation*}

Let us introduce a useful lemma, which is known as the {\em
crypto-lemma} and frequently used in the rest of this paper. The
lemma is given in \cite{ForneyAllerton03}, and we repeat it here for
completeness.

\begin{lemma} [Crypto-lemma \cite{ForneyAllerton03}]
Let $\mathcal{C}$ be a finite or compact group with group operation
$+$. For independent random variables $a$ and $b$ over
$\mathcal{C}$, let $c=a+b$. If $a$ is uniform over $\mathcal{C}$,
then $c$ is independent of $b$ and uniform over $\mathcal{C}$.
\label{LEM:CryptoLemma}
\end{lemma}

By Lemma \ref{LEM:CryptoLemma}, ${\bf X}_i$ is uniformly distributed
over $\mathcal{R}_i$ and independent of ${\bf W}_i$. Thus,
regardless of ${\bf W}_i$, the average transmit power of node $i$ is
equal to $\sigma^2 (\Lambda_i)$, which approaches $P_i$ as $n$ tends
to infinity. Thus, the power constraint is met.

\vspace{2mm}{\bf \em Decoding}\vspace{2mm}

Upon receiving ${\bf Y} = \sum_{j=1}^K {\bf X}_j + {\bf Z}$, where
${\bf Z}$ is a vector of i.i.d. Gaussian noise with zero mean and
unit variance, the receiver computes
\begin{align*}
\tilde{\bf Y} &= \left(\alpha {\bf Y} - \sum_{j=1}^K {\bf
U}_j\right) \md_1 \nonumber\\
&= \Bigg[ \sum_{j=1}^K ({\bf W}_j + {\bf U}_j ) \md_j  -
\sum_{j=1}^K {\bf X}_j \nonumber\\
& \;\;\;+ \alpha \sum_{j=1}^K {\bf X}_j + \alpha {\bf Z} -
\sum_{j=1}^K
{\bf U}_j \Bigg] \md_1 \nonumber\\
&= \left( {\bf T} + \tilde{\bf Z} \right) \md_1 \text,
\end{align*}
where
\begin{align}
{\bf T} &= \left[ \sum_{j=1}^K \left( {\bf W}_j - Q_j (
{\bf W}_j + {\bf U}_j) \right) \right] \md_1 \nonumber\\
&= \left[ {\bf W}_1 + \sum_{j=2}^K \left( {\bf W}_j - Q_j (
{\bf W}_j + {\bf U}_j) \right) \right] \md_1 \text, \label{EQ:DetMAC}\\
\tilde{\bf Z} &= - (1- \alpha) \sum_{j=1}^K {\bf X}_j + \alpha {\bf
Z} \nonumber \text,
\end{align}
$0 \leq \alpha \leq 1$ is a scaling factor, and $Q_j (\cdot)$
denotes the nearest neighbor lattice quantizer associated with
$\Lambda_j$. We choose $\alpha$ as the minimum mean-square error
(MMSE) coefficient to minimize the variance of the effective noise
$\tilde{\bf Z}$. Thus,
\begin{equation*}
\alpha = \frac{\sum_{j=1}^K P_j} {\sum_{j=1}^K P_j + 1} \text,
\end{equation*}
and the resulting noise variance satisfies
\begin{equation}
\frac{1}{n} E\left\{ \left\| \tilde{\bf Z} \right\|^2 \right\} \leq
\frac{\sum_{j=1}^K P_j} {\sum_{j=1}^K P_j + 1} \text.
\label{EQ:NoiseVar}
\end{equation}
Note that, though the relation in (\ref{EQ:NoiseVar}) is given by an
inequality, it becomes tight as $n \rightarrow \infty$ by Theorem
\ref{TH:LatticeChain}. By the chain relation of the lattices in
Theorem \ref{TH:LatticeChain}, it is easy to show that ${\bf T} \in
\mathcal{C}_1$. Regarding ${\bf T}$, we have the following lemma.

\begin{lemma}
${\bf T}$ is uniform over $\mathcal{C}_1$ and independent of
$\tilde{\bf Z}$. \label{LEM:IndepEffNoise}
\end{lemma}

\begin{IEEEproof}
Define $\tilde{\bf W} \triangleq \sum_{j=2}^K \left( {\bf W}_j - Q_j
( {\bf W}_j + {\bf U}_j) \right) \md_1$, and, thus, ${\bf T} =
\left( {\bf W}_1 + \tilde{\bf W} \right) \md_1$. Note that
$\tilde{\bf W}$ is correlated with ${\bf X}_i$, $2 \leq i \leq K$,
and $\tilde{\bf
Z}$. 
Since ${\bf W}_1$ is uniform over $\mathcal{C}_1$ and independent of
$\tilde{\bf W}$, ${\bf T}$ is independent of $\tilde{\bf W}$ and
uniformly distributed over $\mathcal{C}_1$ (crypto-lemma). Hence, if
${\bf T}$ and $\tilde{\bf Z}$ are correlated, it is only through
${\bf W}_1$. However, ${\bf W}_1$ and $\tilde{\bf Z}$ are
independent of each other, and, consequently, ${\bf T}$ is also
independent of $\tilde{\bf Z}$.
\end{IEEEproof}

The receiver tries to retrieve ${\bf T}$ from $\tilde{\bf Y}$
instead of recovering ${\bf W}_i$, $1 \leq i \leq K$, separately.
For the decoding method, we consider {\em Euclidean lattice
decoding} \cite{ErezIT04}-\cite{ForneyAllerton03}, which finds the
closest point to $\tilde{\bf Y}$ in $\Lambda_C$. From the symmetry
of the lattice structure and the independence between ${\bf T}$ and
$\tilde{\bf Z}$ (Lemma \ref{LEM:IndepEffNoise}), the probability of
decoding error is given by
\begin{align}
p_e &= {\rm Pr} \left\{ {\bf T} \neq Q_C \left( \tilde{\bf Y}
\right) \right\} \nonumber\\
&= {\rm Pr} \left\{ \tilde{\bf Z} \md_1 \notin \mathcal{R}_C
\right\} \text, \label{EQ:ErrorProb}
\end{align}
where $Q_C(\cdot)$ denotes the nearest neighbor lattice quantizer
associated with $\Lambda_C$ and $\mathcal{R}_C$ denotes the Voronoi
region of $\Lambda_C$. Then, we have the following theorem.

\begin{theorem}
Let
\begin{equation*}
R_1^* = \left[ \frac{1}{2} \log \left( \frac{P_1}{\sum_{j=1}^K P_j}
+ P_1 \right) \right]^+ \text.
\end{equation*}
For any $\bar{R}_1 < R_1^*$ and a lattice chain as described in
Theorem \ref{TH:LatticeChain} with $R_1$ approaching $\bar{R}_1$,
i.e., $R_1 = \bar{R}_1 + o_n(1)$, the error probability under
Euclidean lattice decoding (\ref{EQ:ErrorProb}) is bounded by
\begin{equation*}
p_e \leq e^{-n \left( E_P \left( 2^{2(R_1^* - \bar{R}_1)} \right) -
o_n(1) \right)} \text.
\end{equation*}
\label{TH:LatticeAch}
\end{theorem}

\begin{proof}
See Appendix \ref{SEC:LatticeAch}.
\end{proof}

According to Theorem \ref{TH:LatticeAch}, the error probability
vanishes as $n \rightarrow \infty$ if $\bar{R}_1 < R_1^*$ since $E_p
(x)
>0$ for $x>1$. 
This implies that the nested lattice code can achieve any rate below
$R_1^*$. Thus, by c) of Theorem \ref{TH:LatticeChain} and Theorem
\ref{TH:LatticeAch}, the coding rate $R_i$, $1 \leq i \leq K$, can
approach $R_i^*$ arbitrarily closely while keeping $p_e$ arbitrarily
small for sufficiently large $n$, where
\begin{equation}
R_i^* = \left[ \frac{1}{2} \log \left( \frac{P_i}{\sum_{j=1}^K P_j}
+ P_i \right) \right]^+ \text. \label{EQ:CodeRate}
\end{equation}

\begin{remark}
In theorem \ref{TH:LatticeAch}, we showed the error exponent of
lattice decoding and the achievability of $R_1$ 
directly followed. However, if we are only interested in finding the
achievability of $R_1$, not in the error exponent, we can use the
argument on the bounding behavior of lattice decoding in
\cite{LoeligerIT97}, which gives the same result in a much simpler
way.
\end{remark}

\begin{remark}
Since $P_1 \geq \cdots \geq P_K$, we have $R_1^* \geq \cdots \geq
R_K^*$. Now, consider the case that, for some $\hat{i} < K$, the
rates $R_i^*$, $\hat{i}+1 \leq i \leq K$, are zero while $R_i^*$, $1
\leq i \leq \hat{i}$, are nonzero. In this situation, nodes
$\hat{i}+1$, \ldots, $K$ cannot transmit any useful information to
the receiver, and, thus, we can turn them off 
so as not to hinder the transmissions of nodes $1$, \ldots,
$\hat{i}$. Then, the variance of $\tilde{\bf Z}$ decreases and we
have extended rates given by
\begin{equation*}
R_i^* = \left[ \frac{1}{2} \log \left(
\frac{P_i}{\sum_{j=1}^{\hat{i}} P_j} + P_i \right) \right]^+, \; 1
\leq i \leq \hat{i} \text.
\end{equation*}
However, for the ease of exposition, we do not consider the
transmitter turning-off 
technique and assume that nodes $\hat{i}+1$, \ldots, $K$ just
transmit ${\bf X}_i = {\bf U}_i$ when their coding rates are zero.
\end{remark}

\subsection{Achievable multicast rate} \label{SEC:GaussianAch}

We consider $B$ blocks of transmissions from the source to
destinations. Each block consists of $n$ channel uses. In block $k
\in \{1, \ldots, B\}$, an independent and uniform message $W[k] \in
\{1, \ldots, 2^{nR}\}$ is sent from the source node $s$. It takes at
most $L \triangleq B+|V|-2$ blocks for all the $B$ messages to be
received by destination nodes. After receiving $L$ blocks,
destination nodes decode the source message $W \triangleq \left(
W[1], \ldots, W[B] \right)$. Thus, the overall rate is $\frac{B}{L}
R$, which can be arbitrarily close to $R$ by choosing $B$
sufficiently large.

\psfull
\begin{figure} [t]
\begin{center}
\epsfig{file=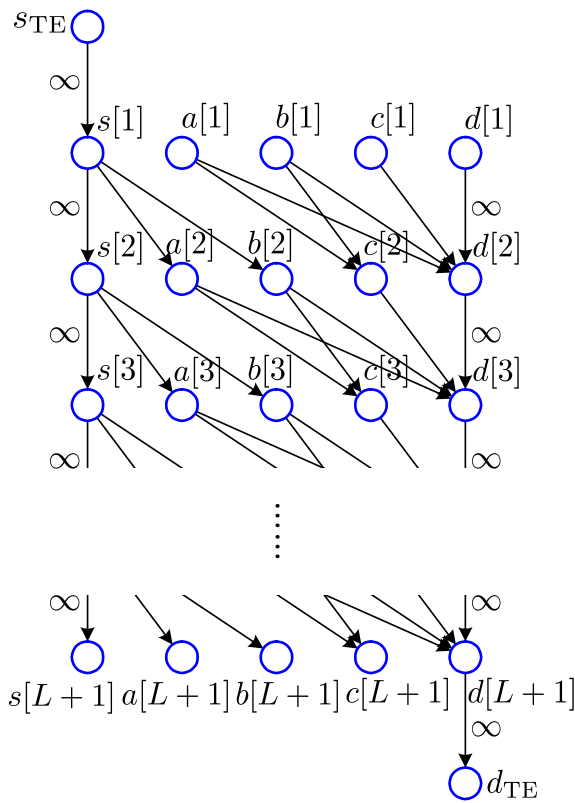, width=3in} \caption{Time-expansion of
the network in Fig. \ref{FIG:General}.} \label{FIG:ExpandGraph}
\end{center}
\end{figure}
\psdraft

\vspace{2mm}{\bf \em Time-expanded network}\vspace{2mm}

For ease of analysis, we consider the $B$ blocks of transmissions
over the time-expanded network \cite{AhlswedeIT00,
AvestimehrAllerton07}, $\mathcal{G}_{\TE}$, obtained by unfolding
the original network $\mathcal{G}$ over $L+1$ time stages. In
$\mathcal{G}_{\TE}$, node $v \in V$ at block $k$ appears as $v[k]$,
and $v[k]$ and $v[k']$ are treated as different nodes if $k \neq
k'$. There are a virtual source and destination nodes, denoted by
$s_{\TE}$ and $d_{\TE}$, respectively. We assume that $s_{\TE}$ and
$s[k]$'s are connected through virtual error-free infinite-capacity
links, and, similarly, $d_{\TE}$ and $d[k]$'s are. For instance, the
network in Fig. \ref{FIG:General} is expanded to the network in Fig.
\ref{FIG:ExpandGraph}. Dealing with the time-expanded network does
not impose any constraints on the network. Any scheme for the
original network can be interpreted to a scheme for the
time-expanded network and vice-versa. In our case, the transmissions
of $B$ messages $W[k]$, $k=1,\ldots, B$, from $s$ to $d \in D$ over
$\mathcal{G}$ correspond to the transmission of a single message $W$
from $s_{\TE}$ to $d_{\TE} \in D_{\TE}$ over $\mathcal{G}_{\TE}$,
where $D_{\TE}$ denotes the set of virtual destination nodes.

A main characteristic of the time-expanded network is that it is
always a layered network \cite{AvestimehrAllerton07} which has equal
length paths from the source to each destination\footnote{Another
characteristic is that the time-expanded network is always acyclic
\cite{AhlswedeIT00}.}. We define the set of nodes at length $k$ from
the virtual source node as
\begin{equation*}
V_{\TE} [k] = \left\{ v[k]: v \in V \right\}
\end{equation*}
and call it the $k$-th layer. We use the subscript `$_{\TE}$' to
differentiate parameters of $\mathcal{G}$ and $\mathcal{G}_{\TE}$.
The set of nodes and edges of $\mathcal{G}_{\TE}$ are defined as
\begin{align*}
V_{\TE} &= \{s_{\TE}\} \cup D_{\TE} \cup \left(
\underset{k=1}{\overset{L+1}{\cup}}
V_{\TE} [k] \right) \text,\\
E_{\TE} &= \left\{ \left( u[k],v[k+1] \right) : (u,v) \in E,
k=1,\ldots,L \right\}\\
&\;\;\; \cup \left\{ (s[k-1],s[k]): k=1,\ldots,L \right\}\\
&\;\;\; \cup \left\{ (d[k],d[k+1]): k=1,\ldots,L \right\} \text,
\end{align*}
where we define $s[0] = s_{\TE}$ and $d [L+2] = d_{\TE}$. Note that,
since $\mathcal{G}_{\TE}$ is layered, edges only appear between
adjacent layers. From $V_{\TE}$ and $E_{\TE}$, the other parameters,
e.g., $\Delta_{\TE} (\cdot)$, $\Theta_{\TE} (\cdot)$, $S_{\TE}$,
$\bar{S}_{\TE}$, $\Gamma_{TE}$, and $\Delta_{\TE,S} (\cdot)$, are
similarly defined as $\Delta (\cdot)$, $\Theta (\cdot)$, $S$,
$\bar{S}$, $\Gamma$, and $\Delta_S (\cdot)$, respectively.

\vspace{2mm}{\bf \em Encoding}\vspace{2mm}

We apply the nested lattice codes in Section \ref{SEC:LatticeCode}
over the all Gaussian MACs in the network. Thus, node $v[k]$ is
assigned with sets of coset leaders $\mathcal{C}_{v[k],w[k^+]}$,
$w[k^+] \in \Theta_{\TE}(v[k])$, where $k^+ \triangleq k+1$. We do
not change the lattice scheme over blocks, and, thus,
$\mathcal{C}_{v[k],w[k^+]} = \mathcal{C}_{v,w}$

At node $s[k]$, the indices $\left\{ 1, \ldots, 2^{nR} \right\}$ are
uniformly randomly mapped onto vectors in $\mathcal{C}_{s,w}$, $w
\in \Theta (s)$. We define the random mapping as $f_{s[k],w[k^+]}
(\cdot)$. Then, node $s[k]$ receives $W = \left( W[1], \ldots, W[B]
\right)$ from $s[k^-]$ through the error-free link, where $k^-
\triangleq k-1$, and transmits
\begin{equation*}
{\bf W}_{s[k],w[k^+]} = f_{s[k],w[k^+]}(W[k])
\end{equation*}
on channel $(s[k],w[k^+])$ using a random dither vector ${\bf
U}_{s[k],w[k^+]}$. At node $v[k]$ that is not $s[k]$ or $d[k]$, the
received signal is given by
\begin{equation}
\tilde{\bf Y}_{v[k]} = \left( {\bf T}_{v[k]} + \tilde{\bf Z}_{v[k]}
\right) \md_v \text, \label{EQ:MultiEnc1}
\end{equation}
where
\begin{equation}
{\bf T}_{v[k]} = \left[ \sum_{u[k^-] \in \atop \Delta_{\TE} (v[k])}
\left( {\bf W}_{u[k^-],v[k]} - Q_{u,v} \left({\bf W}_{u[k^-],v[k]} +
{\bf U}_{u[k^-],v[k]} \right) \right) \right] \md_v \text,
\label{EQ:MultiEnc2}
\end{equation}
and $\tilde{\bf Z}_{v[k]}$ is an effective noise vector. In
(\ref{EQ:MultiEnc1}), $\Lambda_v$ denotes the lattice associated
with the incoming channel to node $v$ with the largest power. Then,
${\bf T}_{v[k]}$ is decoded using Euclidean lattice decoding, which
yields an estimate $\hat{\bf T}_{v[k]}$. Next, $\hat{\bf T}_{v[k]}$
is uniformly and randomly mapped onto vectors in $C_{v,w}$, $w \in
\Theta(v)$. This mapping is denoted by $f_{v[k],w[k^+]} (\cdot)$,
and node $v[k]$ transmits
\begin{equation*}
{\bf W}_{v[k],w[k^+]} = f_{v[k],w[k^+]} \left( \hat{\bf T}_{v[k]}
\right)
\end{equation*}
on channel $(v[k],w[k^+])$ using a random dither vector ${\bf
U}_{v[k],w[k^+]}$. Node $d[k]$, $d \in D$, receives $\tilde{\bf
Y}_{d[k]}$ and computes $\hat{\bf T}_{d[k]}$. It also receives
$\left( \hat{\bf T}_{d[1]}, \ldots, \hat{\bf T}_{d[k^-]} \right)$
from $d[k^-]$ through the virtual error-free infinite-capacity link
and passes $\left( \hat{\bf T}_{d[1]}, \ldots, \hat{\bf T}_{d[k]}
\right)$ to node $d[k^+]$.

We assume that all the random mappings $f_{u[k],v[k^+]}$,
$(u[k],v[k^+]) \in E_{\TE}$ are done independently.

\vspace{2mm}{\bf \em Decoding}\vspace{2mm}

While decoding, a virtual destination node $d_{\TE} \in D_{\TE}$
assumes that there is no error in decoding ${\bf T}_{v[k]}$'s in the
network and that the network is deterministic. Therefore, with
knowledge of all deterministic relations\footnote{It is assumed that
the all random dither vectors are known to destination nodes. Thus,
(\ref{EQ:MultiEnc2}) is deterministic.} (\ref{EQ:MultiEnc2}) in the
network, node $d_{\TE}$ decodes $W$ by simulating all $2^{nBR}$
messages and finding one that yields the received signal $\hat{\bf
T}_{d_{\TE}} \triangleq \left( \hat{\bf T}_{d[1]}, \ldots, \hat{\bf
T}_{d[L+1]} \right)$.

\vspace{2mm}{\bf \em Calculation of the probability of
error}\vspace{2mm}

In the above decoding rule, we will declare an error if at least one
of the following events occurs.
\begin{itemize}
\item $\mathcal{E}_1$: there is an error in decoding ${\bf
T}_{v[k]}$ at at least one node in the network.
\item $\mathcal{E}_2$: a message $W' \neq W$ exists that yields the
same received signal $\hat{\bf T}_{d_{\TE}}$, which is obtained
under $W$, at at least one virtual destination node $d_{\TE} \in
D_{\TE}$.
\end{itemize}
Thus, the error probability is given by
\begin{align}
P_e &= \Pr \{ \mathcal{E}_1 \cup \mathcal{E}_2 \} \nonumber\\
&\leq \Pr \{ \mathcal{E}_1 \} + \Pr \{ \mathcal{E}_2 |
\mathcal{E}_1^c \} \text. \label{EQ:MultiErr1}
\end{align}

Let us consider the first term in (\ref{EQ:MultiErr1}). Using the
union bound, we have
\begin{equation*}
\Pr \{ \mathcal{E}_1 \} \leq \sum_{k=2}^{L+1} \sum_{v[k] \in V[k]
\atop \setminus \{s[k]\}} p_{e,v[k]} \text,
\end{equation*}
where
\begin{equation*}
p_{e,v[k]} \triangleq \Pr \left\{ \hat{\bf T}_{v[k]} \neq {\bf
T}_{v[k]} \right\} \text.
\end{equation*}
Note that the summation is from $k=2$ since nodes in the first layer
do not have any received signal except for node $s[1]$. By Theorem
\ref{TH:LatticeAch}, at node $v \in V \setminus \{1\}$ for any
$\epsilon
>0$, $p_{e,v[k]}$ is less than $\frac{\epsilon}{2L|V|}$ for
sufficiently large $n$ if
\begin{align}
R_{u,v} &= \frac{1}{n} \log |\mathcal{C}_{u,v}| \nonumber\\
&= \left[ \frac{1}{2} \log \left( \left( \frac{1}{\sum_{u' \in \atop
\Delta(v)} P_{u',v}} + 1 \right) \cdot P_{u,v} \right) - \epsilon
\right]^+ \label{EQ:MultiErr1.5}
\end{align}
for all $u \in \Delta (v)$. Therefore, in this case
\begin{equation*}
\Pr \{ \mathcal{E}_1 \} \leq \frac{\epsilon}{2} \text.
\end{equation*}

Now, we consider the second term in (\ref{EQ:MultiErr1}). Under the
condition $\mathcal{E}_1^c$, we have $\hat{\bf T}_{v[k]} = {\bf
T}_{v[k]}$, and, thus, the network is deterministic. Let us use the
notation ${\bf W}_{u[k^-],v[k]} (W)$ and ${\bf T}_{v[k]} (W)$ to
explicitly denote the signals under message $W$. We say that node
$v[k]$ can distinguish $W$ and $W'$ if ${\bf T}_{v[k]}(W) \neq {\bf
T}_{v[k]} (W')$. Thus, from the argument of a deterministic network
in \cite{AvestimehrAllerton07}, the error probability is bounded by
\begin{align}
\Pr \{ \mathcal{E}_2 | \mathcal{E}_1^c \} &\leq 2^{nBR} \cdot \Pr
\left\{ \underset{d_{\TE} \in \atop D_{\TE}}{\cup} \left\{ {\bf
T}_{d_{\TE}} (W) = {\bf T}_{d_{\TE}} (W') \right\} \right\} \nonumber\\
&= 2^{nBR} \cdot \sum_{S_{\TE} \in \atop \Gamma_{\TE}} \Pr \{
\text{\footnotesize Nodes in $S_{\TE}$ can distinguish $W$, $W'$,
and nodes in $S_{\TE}^c$ cannot} \} \text. \label{EQ:MultiErr2}
\end{align}
We briefly denote the probabilities in the summation in
(\ref{EQ:MultiErr2}) as
\begin{equation*}
\Pr \left\{ \mathcal{D} = S_{\TE}, \bar{\mathcal{D}} = S_{\TE}^c
\right\} \text.
\end{equation*}

Here, we redefine the cut in the time-expanded network
$\mathcal{G}_{\TE}$ for convenience sake. From the encoding scheme,
since the source message propagates through nodes $s[k]$, $k=1,
\ldots, L+1$, they can clearly distinguish $W$ and $W'$. Similarly,
if a virtual destination node $d_{\TE}$ cannot distinguish $W$ and
$W'$, nodes $d[k]$, $k=1, \ldots, L+1$ cannot either. Thus, when we
analyze the error probability (\ref{EQ:MultiErr2}), we can always
assume that $s[k] \in S_{\TE}$ and $d[k] \in S_{\TE}^c$, $k=1,
\ldots, L+1$, without loss of generality.

From the fact that $\mathcal{G}_{\TE}$ is layered, we have
\begin{align}
\Pr \left\{ \mathcal{D}= S_{\TE}, \bar{\mathcal{D}}= S_{\TE}^c
\right\} &= \Pr \left\{ \mathcal{D}=S_{\TE}, \bar{\mathcal{D}} =
S_{\TE}^c[1] \right\} \nonumber\\
&\;\; \cdot \prod_{k=2}^{L+1} \Pr \left\{ \bar{\mathcal{D}} =
S_{\TE}^c [k] | \mathcal{D}=S_{\TE} [k^-],
\bar{\mathcal{D}} = S_{\TE}^c [k^-] \right\} \nonumber\\
&\leq \prod_{k=2}^{L+1} \Pr \left\{ \bar{\mathcal{D}} = S_{\TE}^c
[k] | \mathcal{D}=S_{\TE} [k^-], \bar{\mathcal{D}} = S_{\TE}^c [k^-]
\right\} \text, \label{EQ:MultiErr3}
\end{align}
where $S_{\TE} [k]$ and $S_{\TE}^c [k]$ denote the sets of nodes in
$S_{\TE}$ and $S_{\TE}^c$ in the $k$-th layer, i.e.,
\begin{align*}
S_{\TE} [k] &\triangleq S_{\TE} \cap V_{\TE}[k] \text,\\
S_{\TE}^c [k] &\triangleq S_{\TE}^c \cap V_{\TE}[k] \text.
\end{align*}
Also, from the fact that the random mapping for each channel is
independent, we have
\begin{gather}
\Pr \left\{ \bar{\mathcal{D}} = S_{\TE}^c [k] | \mathcal{D}=S_{\TE}
[k^-], \bar{\mathcal{D}} = S_{\TE}^c [k^-] \right\} \nonumber\\
= \prod_{v[k] \in \atop S_{\TE}^c [k]} \Pr \left\{ \bar{\mathcal{D}}
= \{v[k] \} | \mathcal{D}=S_{\TE} [k^-], \bar{\mathcal{D}} =
S_{\TE}^c [k^-] \right\} \text.  \label{EQ:MultiErr4}
\end{gather}
Then, we have the following lemma.

\begin{lemma}
Consider the time-expanded network $\mathcal{G}_{\TE}$ with
independent uniform random mapping at each node. For any
cut\footnote{From the definition, $s[k] \in S_{\TE}$ and $d[k] \in
S_{\TE}^c$, $k=1, \ldots, L+1$.} $S_{\TE}$ in $\mathcal{G}_{\TE}$,
we have
\begin{equation*}
\Pr \left\{ \bar{\mathcal{D}} = \{v[k] \} | \mathcal{D}=S_{\TE}
[k^-], \bar{\mathcal{D}} = S_{\TE}^c [k^-] \right\} \leq 2^{-n
\left( \underset{u[k^-] \in \atop \Delta_{\TE,S} (v[k])}{\max}
R_{u,v}\right) }
\end{equation*}
for node $v[k] \in \bar{S}_{\TE}^c [k]$, where $\bar{S}_{\TE}^c [k]
\triangleq \bar{S}_{\TE}^c \cap V_{\TE}[k]$. For node $v[k] \in
S_{\TE}^c [k] \setminus \bar{S}_{\TE}^c [k]$, we have
\begin{equation*}
\Pr \left\{ \bar{\mathcal{D}} = \{v[k] \} | \mathcal{D}=S_{\TE}
[k^-], \bar{\mathcal{D}} = S_{\TE}^c [k^-] \right\} = 1\text.
\end{equation*}
\label{LEM:MultiErr1}
\end{lemma}

\begin{IEEEproof}
See Appendix \ref{SEC:MultiErr1}.
\end{IEEEproof}

Thus, by (\ref{EQ:MultiErr2})-(\ref{EQ:MultiErr4}) and Lemma
\ref{LEM:MultiErr1}, it follows that
\begin{equation}
\Pr \{ \mathcal{E}_2 | \mathcal{E}_1^c \} \leq 2^{nBR} \cdot
|\Gamma_{\TE}| \cdot 2^{ -n \underset{S_{\TE} \in \atop
\Gamma_{\TE}}{\min} \underset{k=2}{\overset{L+1}{\sum}}
\underset{v[k] \in \atop \bar{S}_{\TE}^c [k]}{\sum} \left(
\underset{u[k^-] \in \atop \Delta_{\TE,S} (v[k])}{\max} R_{u,v}
\right) } \text. \label{EQ:MultiErr5}
\end{equation}
We now consider the following lemma.

\begin{lemma}
In the time-expanded $\mathcal{G}_{\TE}$ with $L+1$ layers, the term
in the exponent of (\ref{EQ:MultiErr5})
\begin{equation*}
\underset{S_{\TE} \in \Gamma_{\TE}}{\min} \sum_{k=2}^{L+1}
\sum_{v[k] \in \atop \bar{S}_{\TE}^c [k]} \left( \underset{u[k^-]
\in \atop \Delta_{\TE,S} (v[k])}{\max} R_{u,v} \right)
\end{equation*}
is upper bounded by
\begin{equation*}
L \cdot \underset{S \in \Gamma}{\min} \sum_{v \in \bar{S}^c} \left(
\underset{u \in \Delta_S (v)}{\max} R_{u,v} \right) \text,
\end{equation*}
and lower bounded by
\begin{equation*}
\left( L- |\Gamma| + 2 \right) \cdot \underset{S \in \Gamma}{\min}
\sum_{v \in \bar{S}^c} \left( \underset{u \in \Delta_S (v)}{\max}
R_{u,v} \right) \text.
\end{equation*}
\label{LEM:MultiErr2}
\end{lemma}

\begin{IEEEproof}
See Appendix \ref{SEC:MultiErr2}.
\end{IEEEproof}

Therefore, by (\ref{EQ:MultiErr1.5}), (\ref{EQ:MultiErr5}), and
Lemma \ref{LEM:MultiErr2}, $\Pr \{ \mathcal{E}_2 | \mathcal{E}_1^c
\}$ is less than $\frac{\epsilon}{2}$ for sufficiently large $n$ if
\begin{equation}
R < \frac{L-|\Gamma|+2}{B} \cdot \underset{S \in \Gamma}{\min}
\sum_{v \in \bar{S}^c} \left[ \frac{1}{2} \log \left( \left(
\frac{1}{\sum_{u \in \atop \Delta(v)} P_{u,v}} + 1 \right) \cdot
\underset{u \in \atop \Delta_S (v)}{\max} P_{u,v} \right) - \epsilon
\right]^+ \text. \label{EQ:MultiErr6}
\end{equation}
Thus, the total probability of error (\ref{EQ:MultiErr1}) is less
than $\epsilon$, and the achievability follows from
(\ref{EQ:MultiErr6}).

\subsection{Gap between the upper and lower bounds}

To compute the gap between the upper bound (\ref{EQ:GaussianUB}) and
the achievable rate (\ref{EQ:GaussianAch}), we can rely on the
following lemmas.

\begin{lemma}
Assume that $P_1 \geq \cdots \geq P_K \geq 0$. For any nonempty set
$A \subseteq \{1,\ldots,K\}$ and $l = \min A$, we have
\begin{align*}
&\frac{1}{2} \log \left( 1+ \left( \sum_{j \in A} \sqrt{P_j}
\right)^2 \right) \nonumber\\ - &\left[ \frac{1}{2} \log \left(
\left( \frac{1}{\sum_{j=1}^K P_j} +1 \right) P_l \right) \right]^+
\leq \log K \text.
\end{align*} \label{LEM:Gap1}
\end{lemma}

\begin{lemma}
\begin{gather*}
\min\{a_1,\ldots,a_k\} - \min\{b_1,\ldots,b_k\} \\ \leq
\max\{(a_1-b_1),\ldots,(a_k-b_k)\} \text.
\end{gather*} \label{LEM:Gap2}
\end{lemma}

The proof of Lemma \ref{LEM:Gap1} is given in Appendix
\ref{SEC:Gap}, and the proof of Lemma \ref{LEM:Gap2} is omitted
since it is straightforward. Using Lemmas \ref{LEM:Gap1} and
\ref{LEM:Gap2}, the gap in (\ref{EQ:GaussianGap}) directly follows.

\section{Linear finite-field symmetric networks with interference} \label{SEC:Finite}

\psfull
\begin{figure} [t]
\begin{center}
\epsfig{file=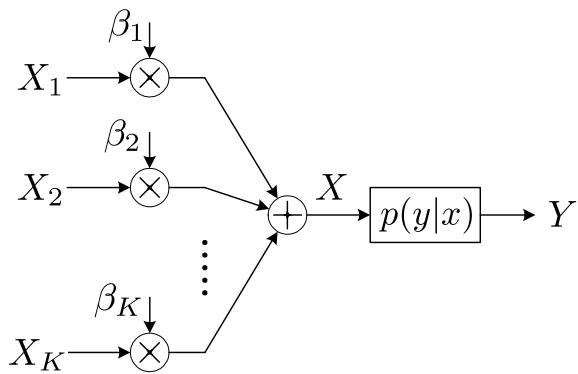, width=3in} \caption{Linear finite-field
symmetric MAC.} \label{FIG:FFSymMAC}
\end{center}
\end{figure}
\psdraft

Let us consider a particular class of discrete memoryless relay
networks with interference. 
The linear finite-field symmetric networks with interference are
characterized by a special structure of MACs in them, which is shown
in Fig. \ref{FIG:FFSymMAC}. In more detail, the linear finite-field
symmetric network with interference is described as follows:
\begin{itemize}
\item Every input alphabet to a MAC at node $v$ is the finite field,
$\mathbb{F}_q$.

\item The received symbol at node $v$, $Y_v^{(t)}$, is determined to be the
output of a {\em symmetric discrete memoryless channel} (DMC)
$\left( \mathbb{F}_q, p(y_v | x_v), \mathcal{Y}_v \right)$ with
input
\begin{equation*}
X_v^{(t)} = \sum_{u \in \Delta(v)} \beta_{u,v} X_{u,v}^{(t)} \text,
\end{equation*}
where $\beta_{u,v} \in \mathbb{F}_q \setminus \{0\}$ denotes the
channel coefficient. For the definition of the symmetric DMC, see
\cite[Sec. 4.5]{GallagerText}.

\item 
The input field size $q$ and channel transition function $p(y_v |
x_v)$ associated with node $v$ need not be identical.
\end{itemize}

A major characteristic of the symmetric DMC is that linear codes can
achieve the capacity \cite[Sec. 6.2]{GallagerText}. Using this,
Nazer and Gastpar \cite{NazerIT07} showed that the {\em computation
capacity} for any linear function of sources can be achieved in the
linear finite-field symmetric MAC in Fig. \ref{FIG:FFSymMAC}. Also,
in \cite{NazerET08, NamAllerton08}, it was shown that linear codes
achieve the multicast capacity of linear finite-field additive noise
and erasure networks with interference, which are special cases of
the class of networks stated above. Extending this line, we
characterize the multicast capacity of the linear finite-field
symmetric network with interference.

\begin{theorem}
The multicast capacity of a linear finite-field symmetric network
with interference is given by
\begin{equation*}
\min_{S \in \Gamma} \sum_{v \in \bar{S}^c} C_v \text,
\end{equation*}
where $C_v$ is the capacity of the channel $\left( \mathbb{F}_q,
p(y_v | x_v), \mathcal{Y}_v \right)$. \label{TH:Finite}
\end{theorem}

The proof of Theorem \ref{TH:Finite} is very similar to the proof of
Theorem \ref{TH:Gaussian}. The difference is that we use linear
codes instead of the nested lattice codes. We show the outline of
the proof in the next subsections.
%

\begin{remark}
The capacity proof for linear finite-field additive noise networks
in \cite{NazerET08} can also be extended to the linear finite-field
symmetric networks in Theorem \ref{TH:Finite}. However, the proof in
\cite{NazerET08} relies on algebraic network coding, and, thus, it
has a restriction on the field size, i.e., $q
> |D|$. In our proof, we do not use the algebraic network coding, and
the field size is not restricted.
\end{remark}

\subsection{Upper bound}

As in the Gaussian case in Section \ref{SEC:GaussianUB}, the upper
bound follows from the relaxed cut-set bound (\ref{EQ:Rcutset}). In
particular, for the linear finite-field symmetric network with
interference, we have the Markov chain relation
$(X_{\bar{S},\bar{S}^c},X_{S^c,V}) \rightarrow X_{\bar{S}^c}
\rightarrow Y_{\bar{S}^c}$, where $X_{\bar{S}^c} = \{ X_v: v \in
\bar{S}^c \}$. Using the data processing inequality, we have
\begin{align*}
I(X_{\bar{S},\bar{S}^c}; Y_{\bar{S}^c}| X_{S^c,V}) &\leq
I(X_{\bar{S}^c}; Y_{\bar{S}^c}| X_{S^c,V})\\
&\leq I(X_{\bar{S}^c}; Y_{\bar{S}^c}) \text.
\end{align*}
Thus the upper bound is given by
\begin{align*}
R &\leq \underset{S \in \Gamma}{\min} \underset{p(x_{V,V})}{\max}
I(X_{\bar{S},\bar{S}^c}; Y_{\bar{S}^c}| X_{S^c,V})\\
&\leq \underset{S \in \Gamma}{\min} \underset{p(x_{V,V})}{\max}
I(X_{\bar{S}^c}; Y_{\bar{S}^c})\\
&= \underset{S \in \Gamma}{\min} \sum_{v \in \bar{S}^c} C_v \text.
\end{align*}

\subsection{Achievability}

Let us denote the vectors of channel input and output of the
symmetric DMC $\left( \mathbb{F}_q, p(y_v | x_v), \mathcal{Y}_v
\right)$ as ${\bf X}_v = \left[ X_v^{(1)},\ldots,X_v^{(n)}
\right]^T$ and ${\bf Y}_v = \left[ Y_v^{(1)},\ldots,Y_v^{(n)}
\right]^T$, respectively. Without loss of generality, we assume that
the encoder input is given by a uniform random vector ${\bf W}_v \in
\mathbb{F}_q^{\lfloor nR_v' \rfloor}$ for some $R_v' \leq 1$. Then
we have the following lemma related to linear coding for the DMC.

\begin{lemma} [Lemma 3 of \cite{NazerIT07}]
For the symmetric DMC $\left( \mathbb{F}_q, p(y_v | x_v),
\mathcal{Y}_v \right)$, a sequence of matrices ${\bf F}_v \in
\mathbb{F}_q^{n \times \lfloor n R_v' \rfloor}$ and associated
decoding function $g_v (\cdot)$ exist such that when ${\bf X}_v =
{\bf F}_v {\bf W}_v$, ${\rm Pr} \{ g({\bf Y}_v) \neq {\bf W}_v \}
\leq \epsilon$ for any $\epsilon >0$ and $n$ large enough if $R_v
\triangleq R_v' \log q < C_v$. \label{LEM:FiniteAch1}
\end{lemma}

We now consider linear encoding for nodes in the network. We let
\begin{equation*}
{\bf X}_{u,v} = \beta_{u,v}^{-1} {\bf F}_v {\bf W}_{u,v},
\end{equation*}
and thus,
\begin{equation*}
{\bf X}_v = \sum_{u \in \Delta(v)} \beta_{u,v} {\bf X}_{u,v} = {\bf
F}_v {\bf T}_v \text,
\end{equation*}
where
\begin{equation}
{\bf T}_v \triangleq \sum_{u \in \Delta(v)} {\bf W}_{u,v} \text.
\label{EQ:FiniteDetRelation}
\end{equation}
By Lemma \ref{LEM:FiniteAch1}, a linear code with sufficiently large
dimension exists such that node $v$ can recover ${\bf T}_v$ with an
arbitrarily small error probability if $R_v < C_v$. Now, we can do
the same as in Section \ref{SEC:GaussianAch} with
(\ref{EQ:FiniteDetRelation}) replacing (\ref{EQ:MultiEnc2}), and the
achievability part follows.

\section{Conclusion} \label{SEC:Conclusion}

In this paper, we considered the multicast problem for relay
networks with interference and examined roles of some structured
codes for the networks. Initially, we showed that nested lattice
codes can achieve the multicast capacity of Gaussian relay networks
with interference within a constant gap determined by the network
topology. We also showed that linear codes achieve the multicast
capacity of linear finite-field symmetric networks with
interference. Finally, we should note that this work
is an intermediate step toward more general networks. 
As an extension to multiple source networks, we showed that the same
lattice coding scheme considered in this work can achieve the
capacity of the Gaussian two-way relay channel within $\frac{1}{2}$
bit \cite{NamIZS08, NamTRC}. As another direction of extension, we
can consider applying structured codes to networks with
non-orthogonal broadcast channels. There is a recent work on the
interference channel \cite{Sridharan08} which is related to this
issue.

\appendix
\subsection{Proof of Theorem \ref{TH:LatticeChain}}
\label{SEC:LatticeChain}

Consider a lattice (more precisely, a sequence of lattices)
$\Lambda_1^n$ with $\sigma^2 (\Lambda_1^n) = P_1$, which is
simultaneously Rogers-good and Poltyrev-good (simultaneously good
shortly). In \cite{ErezIT05}, it was shown that such a lattice
always exists. Then, by the argument in \cite{Krithivasan07}, we can
find a fine lattice $\Lambda_2^n$ such that $\Lambda_1^n \subseteq
\Lambda_2^n$ and $\Lambda_2^n$ is also simultaneously good. We let
the partitioning ratio be
\begin{equation}
\left( \frac{{\rm Vol}(\Lambda_1^n)}{{\rm Vol}(\Lambda_2^n)}
\right)^{\frac{1}{n}} = \left( \frac{P_1}{P_2 - \delta'}
\right)^{\frac{1}{2}} \left( \frac{1}{2 \pi e G(\Lambda_1^n)}
\right)^{\frac{1}{2}} \label{EQ:AP1-1}
\end{equation}
for some $\delta' >0$. Since the partitioning ratio can approach an
arbitrary value as $n$ tends to infinity, for any $\delta > 0$, $n'$
exists such that we can choose $\delta' \leq \delta$ when $n \geq
n'$.
We now have
\begin{align*}
\sigma^2 (\Lambda_2^n) &= G(\Lambda_2^n) \cdot {\rm
Vol}(\Lambda_2^n)^{\frac{2}{n}} \\
& = G(\Lambda_2^n) \cdot 2 \pi e (P_2 - \delta') \text,
\end{align*}
where the second equality follows from (\ref{EQ:AP1-1}). Since
$\Lambda_2^n$ is Rogers-good, $n''$ exists such that $1 \leq 2 \pi e
G(\Lambda_2^n) \leq \frac{P_2}{P_2 - \delta'}$, for $n \geq n''$.
Thus, for $n \geq \max \{ n', n'' \}$, we have
\begin{equation*}
P_2 - \delta \leq \sigma^2 (\Lambda_2^n) \leq P_2 \text.
\end{equation*}
By repeating the same procedure, we obtain a lattice chain
$\Lambda_1^n \subseteq \Lambda_2^n \subseteq \cdots \subseteq
\Lambda_K^n$, where $\Lambda_i^n$, $1 \leq i \leq K$, are
simultaneously good and $P_i - \delta \leq \sigma^2 (\Lambda_i^n)
\leq P_i$ for sufficiently large $n$.

Moreover, by Theorem 5 of \cite{ErezIT04}, if $\Lambda_K^n$ is
simultaneously good, a Poltyrev-good lattice $\Lambda_C^n$ exists
such that $\Lambda_K^n \subseteq \Lambda_C^n$ and the coding rate
$R_K$ can be arbitrary as $n \rightarrow \infty$, i.e.,
\begin{equation*}
R_K = \frac{1}{n} \log \left(
\frac{\vol(\Lambda_K^n)}{\vol(\Lambda_C^n)} \right) = \gamma +
o_n(1)\text.
\end{equation*}
Given $R_K$, the coding rates $R_i$, $1 \leq i \leq K-1$, are given
by
\begin{align*}
R_i &= \frac{1}{n} \log \left( \frac{{\rm Vol}(\Lambda_i^n)}{{\rm
Vol}(\Lambda_C^n)} \right) \\
&= \frac{1}{n} \log \left( \frac{{\rm Vol}(\Lambda_i^n)}{{\rm
Vol}(\Lambda_K^n)} \right) + R_K \\
&= \frac{1}{2} \log \left(
\frac{\sigma^2(\Lambda_i^n)}{\sigma^2(\Lambda_K^n)} \right) + R_K +
o_n(1)\\
&= \frac{1}{2} \log \left( \frac{P_i}{P_K} \right) + R_K + o_n(1)
\text,
\end{align*}
where the third equality follows by the fact that $\Lambda_i^n$ and
$\Lambda_K^n$ are both Rogers-good, and the fourth follows by the
fact that $\sigma^2 (\Lambda_i^n) = P_i - o_n(1)$. \hfill$\Box$

\subsection{Proof of Theorem \ref{TH:LatticeAch}}
\label{SEC:LatticeAch}

Let $r_i^{\cov}$ and $r_i^{\eff}$ denote the covering and effective
radii of $\Lambda_i$, respectively. Then the second moment per
dimension of $r_i^{\cov} \mathcal{B}$ is given by
\begin{equation*}
\sigma_i^2 \triangleq \sigma^2 ( r_i^{\cov} \mathcal{B} ) =
\frac{(r_i^{\cov})^2}{n+2} \text.
\end{equation*}
Next, we define independent Gaussian random variables
\begin{equation*}
{\bf Z}_i \sim \mathcal{N} ( {\bf 0}, \sigma_i^2 {\bf I}), \;
i=1,\ldots,K \text,
\end{equation*}
and
\begin{equation*}
{\bf Z}^* = (1-\alpha) \sum_{j=1}^{K} {\bf Z}_j + \alpha {\bf
Z}\text.
\end{equation*}
Then, we have the following lemmas.

\begin{lemma}
The variance of $Z^*$, each element of ${\bf Z}^*$, is denoted by
${\rm Var}(Z^*)$ and satisfies
\begin{align*}
{\rm Var} (Z^*) &= (1-\alpha)^2 \sum_{j=1}^K \sigma_j^2 + \alpha^2\\
&\leq  \underset{j}{\max} \; \left(\frac{r_j^{\cov}}{r_j^{\eff}}
\right)^2 \cdot \frac{\sum_{j=1}^K P_j}{\sum_{j=1}^K P_j +1} \text.
\end{align*}
\label{LEM:Exponent1}
\end{lemma}


\begin{lemma}
The pdf of $\tilde{\bf Z}$, denoted by $p_{\tilde{\bf Z}}({\bf x})$
satisfies
\begin{equation*}
p_{\tilde{\bf Z}}({\bf x}) \leq e^{n \sum_{j=1}^K \epsilon_j} \cdot
p_{{\bf Z}^*}({\bf x}) \text,
\end{equation*}
where
\begin{equation*}
\epsilon_j = \log \left(\frac{r_j^{\cov}}{r_j^{\eff}} \right) +
\frac{1}{2} \log {2 \pi e G(\mathcal{B})} + \frac{1}{n}.
\end{equation*}
\label{LEM:Exponent2}
\end{lemma}


The above two lemmas are slight modifications of Lemmas 6 and 11 in
\cite{ErezIT04}. The proofs also follow immediately from
\cite{ErezIT04}.

Now, we bound the error probability by
\begin{align}
p_e &= {\rm Pr} \left\{ \tilde{\bf Z} \md_1 \notin \mathcal{R}_C
\right\} \nonumber\\
&\leq {\rm Pr} \left\{ \tilde{\bf Z} \notin \mathcal{R}_C \right\}
\nonumber\\
&\leq e^{n \sum_{j=1}^K \epsilon_j} \cdot {\rm Pr} \left\{ {\bf Z}^*
\notin \mathcal{R}_C \right\} \text, \label{EQ:PfError}
\end{align}
where (\ref{EQ:PfError}) follows from Lemma \ref{LEM:Exponent2}.
Note that ${\bf Z}^*$ is a vector of i.i.d. zero-mean Gaussian
random variables, and the VNR of $\Lambda_C$ relative to ${\bf Z}^*$
is given by
\begin{align}
\mu &= \frac{(\vol(\Lambda_C))^{2/n}}{2 \pi e {\rm Var}(Z^*)}
\nonumber\\
&\geq \frac{(\vol(\Lambda_1))^{2/n}/2^{2R_1}}{2 \pi e
\cdot \frac{\sum_{j=1}^K P_j}{\sum_{j=1}^K P_j +1}} - o_n (1) \label{EQ:VNR1}\\
&= \frac{1}{2^{2R_1}} \cdot \frac{1}{2 \pi e G(\Lambda_1)} \cdot
\left( \frac{P_1}{\sum_{j=1}^K P_j} + P_1 \right) - o_n(1) \label{EQ:VNR2}\\
&= \frac{1}{2^{2 \bar{R}_1}} \cdot \left( \frac{P_1}{\sum_{j=1}^K
P_j} + P_1 \right) - o_n(1) \text, \label{EQ:VNR3}
\end{align}
where (\ref{EQ:VNR1}) follows from Lemma \ref{LEM:Exponent1} and the
fact that $\Lambda_i$, $1\leq i \leq K$, are Rogers-good,
(\ref{EQ:VNR2}) from the definition of $G(\Lambda_1)$, and
(\ref{EQ:VNR3}) from the fact that $\Lambda_1$ is Rogers-good and
$R_1 = \bar{R}_1 + o_n (1)$. When we consider the Poltyrev exponent,
we are only interested in the case that $\mu > 1$. Thus, from the
definition of $R_1^*$ and (\ref{EQ:VNR3}), we can write
\begin{equation*}
\mu = 2^{2(R_1^* - \bar{R}_1)} - o_n(1) \text,
\end{equation*}
for $\bar{R}_1 < R_1^*$. Finally, from (\ref{EQ:PfError}) and by the
fact that $\Lambda_C$ is Poltyrev-good, we have
\begin{align*}
p_e &\leq e^{n \sum_{j=1}^K \epsilon_j} \cdot e^{-n E_P(\mu)} \\
& = e^{-n \left( E_P \left( 2^{2(R_1^* - \bar{R}_1)} \right) -
o_n(1) \right)} \text.
\end{align*}
\hfill$\Box$

\subsection{Proof of Lemma \ref{LEM:MultiErr1}} \label{SEC:MultiErr1}


For notational simplicity, we prove this lemma in the standard MAC
in Section \ref{SEC:LatticeCode}. We assume that the uniform random
mapping is done at each input node of the standard MAC, as was done
in the network. Let $A$ and $A^c$ be nonempty partitions of
$\{1,\ldots,K\}$, i.e., $A \cup A^c = \{1,\ldots,K\}$, and $A \cap
A^c = \emptyset$. We assume that $A$ implies the set of nodes that
can distinguish $W$ and $W'$, and $A^c$ implies the set of nodes
that cannot. For node $i \in A$, ${\bf W}_i (W)$ and ${\bf W}_i
(W')$ are uniform over $\mathcal{C}_i$ and independent of each other
due to the uniform random mapping. However, for node $i \in A^c$, we
always have ${\bf W}_i (W) = {\bf W}_i (W')$. Thus, if $A =
\emptyset$, ${\bf T} (W) = {\bf T} (W')$ always holds, i.e.,
\begin{equation*}
\Pr \left\{ {\bf T}(W) = {\bf T}(W') | \mathcal{D}=A,
\bar{\mathcal{D}}=A^c \right\} =1 \text.
\end{equation*}
If $A \neq \emptyset$, given $\mathcal{D}=A$ and $\bar{\mathcal{D}}=
A^c$, the event ${\bf T}(W) = {\bf T}(W')$ is equivalent to
$\tilde{\bf T}(W) = \tilde{\bf T}(W')$, where
\begin{equation*}
\tilde{\bf T}(W) = \left[ \sum_{j \in A} \left( {\bf W}_j (W) -Q_j (
{\bf W}_j (W) + {\bf U}_j) \right) \right] \md_1 \text,
\end{equation*}
and $\tilde{\bf T} (W')$ is given accordingly. Now, let $l
\triangleq \min A$, then
\begin{align*}
{\bf T}' (W) &\triangleq \tilde{\bf T} (W) \md_l \\
&= \left[ {\bf W}_l (W) + \sum_{j \in A \atop \setminus \{l\}}
\left( {\bf W}_j (W) -Q_j ( {\bf W}_j (W) + {\bf U}_j) \right)
\right] \md_l \text,
\end{align*}
which follows from the fact that $\Lambda_1 \subseteq \Lambda_l$,
and ,thus, $({\bf x} \md_1)\md_l = {\bf x} \md_l$. Note that, due to
the crypto-lemma and the uniform random mapping, ${\bf T}' (W)$ and
${\bf T}' (W')$ are uniform over $\mathcal{C}_l$ and independent of
each other. Therefore,
\begin{align*}
\Pr \left\{ {\bf T}(W) = {\bf T}(W') | \mathcal{D}=A,
\bar{\mathcal{D}}=A^c \right\} &= \Pr \left\{ \tilde{\bf T} (W) =
\tilde{\bf T} (W') | \mathcal{D}=A \right\} \\
&\leq \Pr \left\{ {\bf T}' (W) = {\bf T}' (W') | \mathcal{D}=A
\right\} \\
&= \frac{1}{|\mathcal{C}_l|} = 2^{-nR_l} \text.
\end{align*}
Thus, by changing notations properly to those of the network, we
complete the proof. \hfill$\Box$

\subsection{Proof of Lemma \ref{LEM:MultiErr2}}
\label{SEC:MultiErr2}


In the time-expanded network, there are two types of cuts, steady
cuts and wiggling cuts \cite{AvestimehrAllerton07}. The steady cut
separates the nodes in different layers identically. That is, for a
steady cut $S_{\TE}$, $v[k] \in S_{\TE}$ for some $k$ if and only if
$v[1], \ldots, v[L+1] \in S_{\TE}$. Let us denote the set of all
steady cuts as $\tilde{\Gamma}_{\TE}$. Then, since
$\tilde{\Gamma}_{\TE} \subseteq \Gamma_{\TE}$,
\begin{align*}
\underset{S_{\TE} \in \Gamma_{\TE}}{\min} \sum_{k=2}^{L+1}
\sum_{v[k] \in \atop \bar{S}_{\TE}^c [k]} \left( \underset{u[k^-]
\in \atop \Delta_{\TE,S} (v[k])}{\max} R_{u,v} \right) &\leq
\underset{S_{\TE} \in \tilde{\Gamma}_{\TE}}{\min} \sum_{k=2}^{L+1}
\sum_{v[k] \in \atop \bar{S}_{\TE}^c [k]} \left( \underset{u[k^-]
\in \atop
\Delta_{\TE,S} (v[k])}{\max} R_{u,v} \right)\\
&= L \cdot \underset{S \in \Gamma}{\min} \sum_{v \in \bar{S}^c}
\left( \underset{u \in \Delta_S (v)}{\max} R_{u,v} \right) \text.
\end{align*}

We now prove the lower bound. 
For any two cuts $S_1$ and
$S_2$ in $\mathcal{G}$, i.e., $S_1, S_2 \in \Gamma$, define that
\begin{equation*}
\xi (S_1,S_2) = \sum_{v \in S_2^c} \left( \underset{u \in
S_1}{\max}\; R_{u,v} \right) \text,
\end{equation*}
where $R_{u,v} = 0$ if $(u,v) \notin E$. Then, we have the following
lemma

\begin{lemma}
Consider a sequence of non-identical cuts $S_1, \ldots, S_{L'} \in
\Gamma$ and define $S_{L'+1} = S_1$. For the sequence, we have
\begin{equation*}
\sum_{k=1}^{L'} \xi (S_k, S_{k+1}) \geq \sum_{k=1}^{L'} \xi(S_k',
S_k') \text,
\end{equation*}
where for $k=1, \ldots, L'$,
\begin{equation*}
S_k' = \underset{\{i_1,\ldots,i_k\} \subseteq \atop
\{1,\ldots,L'\}}{\cup} (S_{i_1} \cap \cdots \cap S_{i_k}) \text.
\end{equation*}
\label{LEM:Submod}
\end{lemma}

The proof of Lemma \ref{LEM:Submod} is tedious but straightforward.
Similar lemmas were presented and proved in \cite[Lemma
6.4]{AvestimehrAllerton07}, \cite[Lemma 2]{SmithITW07}, and the
proof of Lemma \ref{LEM:Submod} also follows similarly.

Now, since $S_k' \in \Gamma$, it follows that
\begin{align}
\xi(S_k', S_k') &\geq \underset{S \in \Gamma}{\min} \sum_{v \in S^c}
\left( \underset{u \in S}{\max} \; R_{u,v} \right) \nonumber\\
&= \underset{S \in \Gamma}{\min} \sum_{v \in \bar{S}^c} \left(
\underset{u \in \atop \Delta_S (v)}{\max}\: R_{u,v} \right) \text.
\label{EQ:Submod}
\end{align}
Also, since $S_{\TE}[k]$'s correspond to cuts in $V$, we can rewrite
\begin{equation*}
\underset{S_{\TE} \in \Gamma_{\TE}}{\min} \sum_{k=2}^{L+1}
\sum_{v[k] \in \atop \bar{S}_{\TE}^c [k]} \left( \underset{u[k^-]
\in \atop \Delta_{\TE,S} (v[k])}{\max} R_{u,v} \right) =
\underset{S_{\TE} \in \Gamma_{\TE}}{\min} \sum_{k=2}^{L+1} \xi
\left( S_{\TE}[k^-], S_{\TE}[k] \right) \text.
\end{equation*}
Since there are $|\Gamma| = 2^{|V|-2}$ different cuts, at least the
first $L - |\Gamma| +2$ of the sequence $S_{\TE}[1], \ldots,
S_{\TE}[L+1]$ form loops, and, thus, by Lemma \ref{LEM:Submod} and
(\ref{EQ:Submod}), we have
\begin{equation*}
\underset{S_{\TE} \in \Gamma_{\TE}}{\min} \sum_{k=2}^{L+1} \xi
\left( S_{\TE}[k^-], S_{\TE}[k] \right) \geq \left( L - |\Gamma| + 2
\right) \cdot \underset{S \in \Gamma}{\min} \sum_{v \in \bar{S}^c}
\left( \underset{u \in \atop \Delta_S (v)}{\max} R_{u,v}
\right)\text.
\end{equation*}
%
%
\hfill$\Box$

\subsection{Proof of Lemma \ref{LEM:Gap1}} \label{SEC:Gap}

We first consider the case that $1 \in A$, and the case that $1
\notin A$ afterward.

\vspace{2mm}{a) $1 \in A$}\vspace{2mm}

In this case, $l=1$, and the gap is
\begin{align*}
&\frac{1}{2} \log \left( 1+ \left( \sum_{j \in A} \sqrt{P_j}
\right)^2 \right) - \left[ \frac{1}{2} \log \left( \left(
\frac{1}{\sum_{j=1}^K P_j} +1 \right) P_1 \right) \right]^+ \\
&\leq \frac{1}{2} \log \left( 1+ \left( \sum_{j=1}^K \sqrt{P_j}
\right)^2 \right) - \frac{1}{2} \log \left( \left(
\frac{1}{\sum_{j=1}^K P_j} +1 \right) P_1 \right) \\
&\leq \frac{1}{2} \log \left( 1 + K^2 P_1 \right) - \frac{1}{2} \log
\left(\frac{1}{K} + P_1 \right) \\
&\leq \log K \text.
\end{align*}

\vspace{2mm}{b) $1 \notin A$}\vspace{2mm}

Since $1 \notin A$, $|A| \leq K-1$. 
Now, the gap is given by
\begin{align*}
&\frac{1}{2} \log \left( 1+ \left( \sum_{j \in A} \sqrt{P_j}
\right)^2 \right) - \left[ \frac{1}{2} \log \left( \left(
\frac{1}{\sum_{j=1}^K P_j} +1 \right) P_l \right) \right]^+ \\
&\leq \frac{1}{2} \log \left( 1+ (K-1)^2 P_l \right) - \left[
\frac{1}{2} \log P_l \right]^+ \\
&\leq \frac{1}{2} \log (1 + (K-1)^2) \\
&\leq \log K \text.
\end{align*}
\hfill$\Box$


\begin{thebibliography}{1}

\bibitem{CoverIT79}
{T. M. Cover and A. A. El Gamal},
\newblock ``{Capacity theorems for the relay channels},''
\newblock {\em IEEE Trans. Inform. Theory}, vol. 51, no. 5, pp.
572--584, Sep. 1979.

\bibitem{AhlswedeIT00}
{R. Ahlswede, N. Cai, S.-Y. R. Li, and R. W. Yeung},
\newblock ``{Network information flow},''
\newblock {\em IEEE Trans. Inform. Theory}, vol. 46, no. 4, pp.
1204--1216, Oct. 2000.


\bibitem{Aref}
{M. R. Aref},
\newblock ``{Information flow in relay networks},''
Ph.D. dissertation Stanford Univ., Stanford, CA, 1980.

\bibitem{RatnakarIT06}
{N. Ratnakar and G. Kramer},
\newblock ``{The multicast capacity of deterministic relay networks with no interference},''
\newblock {\em IEEE Trans. Inform. Theory}, vol. 52, no. 6, pp. 2425--2432, June 2006.

\bibitem{AvestimehrAllerton07}
{A. S. Avestimehr, S. N. Diggavi, and D. N. C. Tse},
\newblock ``{Wireless network information flow},''
\newblock in {\em Proc. 45th Annual Allerton Conference}, Sept. 2007.

\bibitem{AvestimehrISIT08}
{------},
\newblock ``{Approximate capacity of {Gaussian} relay networks},''
\newblock in {\em Proc. IEEE International Symp. Inform. Theory}, Toronto, Canada, July 2008.

\bibitem{DanaIT06}
{A. Dana, R. Gowaikar, R. Palanki, B. Hassibi, and M. Effros},
\newblock ``{Capacity of wireless erasure networks},''
\newblock in {\em IEEE Trans. Inform. Theory}, vol. 52, no. 3, pp. 789--804, Mar. 2006.

\bibitem{SmithITW07}
{B. Smith and S. Vishwanath},
\newblock ``{Unicast transmission over multiple access erasure networks: Capacity and duality},''
\newblock in {\em IEEE Information Theory Workshop}, Tahoe city, California, Sept. 2007.

\bibitem{NazerISIT06}
{B. Nazer and M. Gastpar},
\newblock ``{Computing over multiple-access channels with connections to wireless network coding},''
\newblock in {\em Proc. IEEE International Symp. Inform. Theory}, Seattle, USA, July 2006.

\bibitem{NazerAllerton07}
{------},
\newblock ``{Lattice coding increases multicast rates for Gaussian multiple-access networks},''
\newblock in {\em Proc. 45th Annual Allerton Conference}, Sept. 2007.

\bibitem{NazerIT07}
{------},
\newblock ``{Computation over multiple-access channels},''
\newblock {\em IEEE Trans. Inform. Theory}, vol. 53, no. 10, pp. 3498--3516, Oct. 2007.

\bibitem{NazerET08}
{------},
\newblock ``{The case for structured random codes in network capacity theorems},''
\newblock {\em European Trans. Telecomm.: Special Issue on New Directions in
Inform. Theory}, no. 4, vol. 19, pp. 455--474, June 2008.

\bibitem{NamAllerton08}
{W. Nam and S.-Y. Chung},
\newblock ``{Relay networks with orthogonal components},''
\newblock in {\em Proc. 46th Annual Allerton Conference}, Sept. 2008.

\bibitem{ElGamalIT05}
{A. El Gamal and S. Zahedi},
\newblock ``{Capacity of a class of relay channels with orthogonal components},''
\newblock {\em IEEE Trans. Inform. Theory}, vol. 51, no. 5, pp.
1815--1817, May 2005.






\bibitem{NamIZS08}
{W. Nam, S.-Y. Chung, and Y. H. Lee},
\newblock ``{Capacity bounds for two-way relay channels},''
\newblock {\em Proc. Int. Zurich Seminar on Comm.}, Mar. 2008.

\bibitem{NarayananAllerton07}
{K. Narayanan, M. P. Wilson, and A. Sprintson},
\newblock ``{Joint physical layer coding and network coding for bi-directional relaying},''
\newblock in {\em Proc. 45th Annual Allerton Conference}, Sept. 2007.

\bibitem{NamTRC}
{W. Nam, S.-Y. Chung, and Y. H. Lee},
\newblock ``{Capacity of the Gaussian Two-way Relay Channel to within $\frac{1}{2}$ Bit},''
\newblock submitted to {\em IEEE Trans. Inform. Theory}, available at
\url{http://arxiv.org/PS_cache/arxiv/pdf/0902/0902.2438v1.pdf}.

\bibitem{ZamirIT02}
{R. Zamir, S. Shamai, and U. Erez},
\newblock ``{Nested linear/lattice codes for structured multiterminal binning},''
\newblock {\em IEEE Trans. Inform. Theory}, vol. 48, no. 6, pp.
1250--1276, June 2002.

\bibitem{ErezIT04}
{U. Erez and R. Zamir},
\newblock ``{Achieving $\frac{1}{2}\log(1+SNR)$ on the AWGN channel with lattice encoding and decoding},''
\newblock {\em IEEE Trans. Inform. Theory}, vol. 50, no. 10, pp.
2293--2314, Oct. 2004.

\bibitem{ErezIT05}
{U. Erez, S. Litsyn, and R. Zamir},
\newblock ``{Lattices which are good for (almost) everything},''
\newblock {\em IEEE Trans. Inform. Theory}, vol. 51, no. 10, pp.
3401--3416, Oct. 2005.

\bibitem{LoeligerIT97}
{H. A. Loeliger},
\newblock ``{Averaging bounds for lattices and linear codes},''
\newblock {\em IEEE Trans. Inform. Theory}, vol. 43, no. 6, pp.
1767--1773, Nov. 1997.

\bibitem{PoltyrevIT94}
{G. Poltyrev},
\newblock ``{On coding without restrictions for the AWGN channel},''
\newblock {\em IEEE Trans. Inform. Theory}, vol. 40, no. 2, pp.
409--417, Mar. 1994.

\bibitem{ForneyAllerton03}
{G. D. Forney Jr.},
\newblock ``{On the role of MMSE estimation in approaching the information theoretic limits of linear Gaussian channels: Shannon meets Wiener},''
\newblock in {\em Proc. 41st Annual Allerton Conference}, Oct. 2003.


\bibitem{Krithivasan07}
{D. Krithivasan and S. S. Pradhan},
\newblock ``{A proof of the existence of good nested lattices},''
\newblock available at \url{http://www.eecs.umich.edu/techreports/systems/cspl/cspl-384.pdf}.


\bibitem{PhilosofISIT07}
{T. Philosof, A. Khisti, U. Erez, and R. Zamir},
\newblock ``{Lattice strategies for the dirty multiple access channel},''
\newblock in {\em Proc. IEEE International Symp. Inform. Theory}, Nice, France, June-July 2007.

\bibitem{Sridharan08}
{S. Sridharan, A. Jafarian, S. Vishwanath, S. A. Jafar, and S.
Shamai},
\newblock ``{A layered lattice coding scheme for a class of three user Gaussian interference channels},''
\newblock available at
\url{http://arxiv.org/PS_cache/arxiv/pdf/0809/0809.4316v1.pdf}.

\bibitem{CoverText}
{T. Cover and J. Thomas},
\newblock {\em Elements of Information Theory}, Wiley, New York,
1991.

\bibitem{GallagerText}
{R. Gallager},
\newblock {\em Information Theory and Reliable Communication}, Wiley, New York,
1968.

\end{thebibliography}
\end{document}